\documentclass[conference]{IEEEtran}

\usepackage{multicol}
\usepackage{url}
\usepackage{todonotes}

\begin{document}
%
\title{Adaptive Artificial Intelligence in Games: Issues, Requirements, and a Solution through Behavlets-based General Player Modelling}

\author{\IEEEauthorblockN{Benjamin Ultan Cowley}
\IEEEauthorblockA{BrainWork Research Centre\\
Finnish Institute of Occupational Health\\
POBox 40, Helsinki 00250, Finland\\
\\
Cognitive Brain Research Group\\
University of Helsinki, Finland\\
Email: ben.cowley@helsinki.fi}
\and
\IEEEauthorblockN{Darryl Charles}
\IEEEauthorblockA{School of Computing \& Information Engineering\\
University of Ulster\\
Northern Ireland\\
Email: dk.charles@ulster.ac.uk}
}

\maketitle

\begin{abstract}
We present the last in a series of three academic essays which deal with the question of how and why to build a generalized player model. We begin with a proposition: a general model of players requires parameters for the subjective experience of play, including at least: player psychology, game structure, and actions of play. Based on this proposition, we pose three linked research questions, which make incomplete progress toward a generalised player model: 
\textsf{RQ1} \textit{what is a necessary and sufficient foundation to a general player model?}; 
\textsf{RQ2} \textit{can such a foundation improve performance of a computational intelligence-based player model?}; and 
\textsf{RQ3} \textit{can such a player model improve efficacy of adaptive artificial intelligence in games?}

We set out the arguments behind these research questions in each of the three essays, presented as three preprints.

The third essay, in this preprint, presents the argument that adaptive game artificial intelligence will be enhanced by a generalised player model. This is because games are inherently human artefacts which therefore, require some encoding of the human perspective in order to effectively autonomously respond to the individual player. The player model informs the necessary \textit{constraints} on the adaptive artificial intelligence. A generalised player model is not only more efficient than a per-game solution, but also allows comparison between games which makes it a useful tool for studying play in general.

We describe the concept and meaning of an adaptive game. 
We propose requirements for functional adaptive AI, arguing from first principles drawn from the games research literature. We propose solutions to these requirements, chiefly formal specification for a generalised player model. 
Finally, we propose a plan for future work to develop the formal model approach and integrate with our existing 'Behavlets' method for psychologically-derived player modelling: 

Cowley, B., \& Charles, D. (2016). Behavlets: a Method for Practical Player Modelling using Psychology-Based Player Traits and Domain Specific Features. \textit{User Modeling and User-Adapted Interaction}, 26(2), 257-306.
\end{abstract}


\pagenumbering{arabic}

\section{Introduction}
\label{intro}
\begin{center}
{\it Whosoever desires constant success must change his conduct with the times.

Niccolo Machiavelli}
\end{center}

Computer games have the potential to adapt themselves through changes to their difficulty, appearance, story or even rules. Computer games thus offer a unique opportunity for play that is completely tailored to the individual player by adaptive artificial intelligence (AI). Any non-trivial adaptation requires a player model, to encode relevant aspects of player individuality. The concept of a generalised player model extends this, to describe the subjective experience of play in terms of validated constructs, which could include psychology profiles, game design patterns, action patterns, and more. The aim is not to provide a one-size-fits-all model, but to use generally applicable descriptors of play experience. Any such model will still have to be customised to work in a particular game.

In this preprint, we suggest that adaptive game AI will be optimised by a generalised player model. This is because games are inherently human artefacts which, therefore, require some encoding of the human perspective in order to effectively autonomously respond to the individual player. This argument is built on the idea that the player model will guide \textit{constraints} which are necessary to impose on the adaptive AI. A generalised player model brings added benefits in this context. It can be more efficient than creating a novel model for every game. It also allows comparison between games which makes it a useful tool for studying play in general.

A generalised player model requires a foundation of parameters that describe the subjective experience of play. The foundation will draw on established modelling tools, including at least: \textsf{a}) psychology of behaviour; \textsf{b}) general game design; and \textsf{c}) actions in the context of a given game. 
This foundation should also be integrated with the computational intelligence that drives the model.

These arguments imply several research questions. 
In the first preprint in this series, \cite{Cowley2016:pre1}, we discussed how to improve the theoretical validity of such a foundation by meta-analysis. In the second preprint \cite{Cowley2016:pre2}, we described how such a foundation can improve algorithmic performance of a real-time player model.
The final research question we define is: \textsf{RQ3} \textit{can such a player model improve efficacy and viability of the artificial intelligence required to power games which adapt to their players?} 

The aim of this preprint is to discuss why adaptive game AI benefits from a general model of player psychology. We first describe the concept and meaning of an adaptive game, and discuss to which aspects of player psychology the game can in fact adapt. 
We then make the case for our argument from first principles, drawing on the literature of games studies. We propose possible but speculative solutions, including a formal category theoretic basis to a generalised player model. 

Finally, we propose a plan for future work to more to comprehensively address \textsf{RQ3}, by developing the formal model approach and integrating with an existing method for psychologically-derived player modelling, termed 'Behavlets'.
We previously proposed the Behavlets method to build facets \textsf{a}) to \textsf{c}) above into composite features of game-play defined over entire action sequences \cite{Cowley2016behavlet}, and thus model players for, e.g. personality type classification \cite{Cowley2013}.

\section{Background}
\label{sec:back}

\subsection{Adaptivity in games}
Generally speaking, there two types of game: single player and multiplayer. Adaptive algorithms may be used in a multiplayer game for a range of reasons. However, the challenge in these games is due mainly to other human players. In an online multiplayer game, with an interface via network but without camera or microphone, the players do not see each other nor do they have access to traditional social cues for understanding their opponents, e.g. body language. Such a game environment is \textit{stigmergic}, as players do not interact directly but through the shared environment of the game space. They will read the signs left by their opponent, build a 'theory of mind' model around the complete set of actions observed, and classify the other player based on both what they know of the types of player of that game, and natural social recognition skills. This process contributes to the decision making process for how to play the game and is very different from a single player game.

In single player games, the inclusion of good quality, artificially intelligent non player characters (NPCs) can be central to gameplay design and is important for the player experience. Games can be created to be adaptive to the player through changes to NPC behaviour, or by altering other parameters of the game that affect the gameplay. In either case, a player learns to be more effective at playing a game by learning the rules of the game, including how NPCs behave. Unlike real players, NPC behaviour is usually more predictable and typically it is easier to develop strategies to be successful in competitive gameplay. Players expect NPCs to behave consistently. Nevertheless, one of the reasons that people like to play against other real players is that it is often more satisfying. People behave differently from each other, have complex capability profiles, and are motivated by different ways of playing. Players understand that other players are less predictable than NPCs and accept this. 

Dynamic Difficulty Adjustment (DDA) is a popular approach to implement adaptive AI, e.g. \cite{Chanel2011}. It can work, for example, by altering the number of power-ups in a game or by making non-player characters more or less co-operative or competitive. Some of the earliest games to implement DDA systems were \textit{Max Payne} (3D Realms, 2001) and \textit{Prey} (3D Realms, 2006). However, the AI can do much more than control an opposing force. Forms of adaptative AI have been demonstrated in several commercial games through changes to their difficulty \textit{Max Payne} (3D Realms, 2001), adjusting player character attributes \textit{MarioKart} (Nintendo, 1992), appearance \textit{Fable} (Lionhead Studios, 2004), story \textit{Facade} (Mateus \& Stern, 2005), character learning \textit{Black and White} (Lionhead Studios, 2001), and reactive squad tactics  \textit{Fear} (Monolith Productions, 2005). Related research covers several disciplinary areas including Game AI, Computational Intelligence and Games, and Machine Learning in Games. Yannakakis reviewed \cite{Yannakakis2012review} the literature across the research community and categorised key research into three main areas: Player Experience Modelling (PEM), Procedural Content Generation (PCM), and Massive-Scale Game Data Mining (MDM). Our research approach is predominantly in the PEM category. 

We consider a game AI system holistically, of which NPC behaviour is a core aspect, in which the AI has two coordinated systems: $\alpha$) a user model to capture some aspect of player psychology, which then supplies parameters to $\beta$) controller(s) for adjusting some relevant game system(s). The user model $\alpha$ is intended to model relevant data about some area of the player's state, as discussed further below \ref{sec:psych}. The issue of what the AI can control is somewhat out of scope, and thus is only briefly discussed here. In many ways the ultimate adaptive system would be based around a human or team of humans who dynamically adjust the gameplay experience based on player choice. Consider a Dungeon Master in table top role play games or the alleged human guidance of the Big Blue chess playing algorithm. In both cases human  guidance provides additional nuances, e.g. flexibility or experience, to the dynamic adjustment of a game playing experience. A game AI based on these principles could be more effective in tailoring fun experiences for a greater range of players. This is part of the motivation for the approach underlying our research.    

\subsection{Modus operandi}
Adaptive AI should be built around creating a \textit{more engaging} game for the individual player. Obviously game playing AI can be created for the sole purpose of beating the opposing player, such as in Chess or Go playing programs. However we consider this to be a separate class of AI where player psychology is more or less irrelevant.

A standard game is constrained by its ruleset, but an adaptive game has the potential to exceed known constraints, or the known/explored state space. 
Thus we contend that \textit{unconstrained} adaptive AI can violate certain principles of good game design, such as logical consistency and a coherent Magic Circle \cite{Huizinga1949}, by exhibiting emergent behaviour. It follows that adaptive AI must be explicitly constrained to adhere to the prior assumptions of the player, which can be encoded as a player model. 

Further, undesirable and unpredictable game play bugs can emerge from adaptive components, which could be difficult to test exhaustively. Thus adaptive AI requires a certain level of \textit{formal understanding} of the game that is being designed, in order to give designers the tools to build meta-constraints in the abstract level of the game's possibility space. 
The use of formal modelling can also benefit the creation of a generalised player model, as described in an earlier publication \cite{Cowley2016atari}.

\subsection{Interaction modes}
It is useful to briefly describe the varying forms of player-game interaction, as this can affect how adaptive AI could be deployed. Salen \& Zimmerman \cite{Salen2004} illustrated four different modes of interactivity in computer games:
\begin{enumerate}
	\item Cognitive. The psychological, emotional and intellectual interaction with the game.

	\item Functional. Essentially this is the interaction with the game interface and the primary means for accessing the game mechanics.

	\item Explicit. The interaction with the underlying game mechanics – this is the core of the game, the mechanics and the formal rules.

	\item Cultural. This occurs outside the bounds of the game in the form of fan sites, creation of and use of cheats etc.  
\end{enumerate}

\begin{figure*}[!ht]
	\centering
	\includegraphics[width = \textwidth]{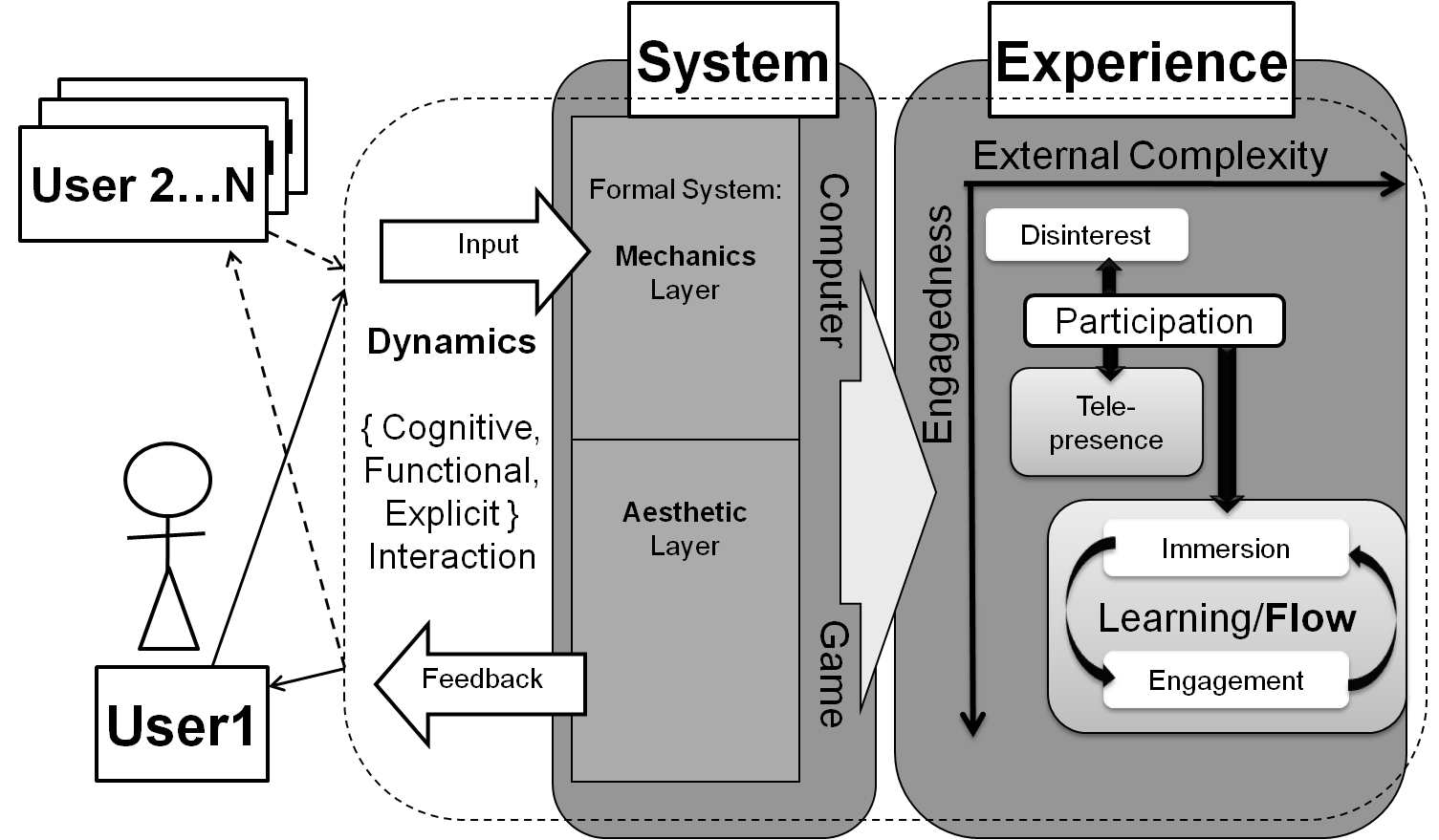}
	\caption{A model of player interaction with a computer game, for two or more players.}
	\label{fig.2pUSE}
\end{figure*}

Here we are interested in the first three modes of interactivity and how these relate to game adaptation. This interactivity is illustrated by the relational schema shown in figure \ref{fig.2pUSE}. This is a combination of ideas following on from LeBlanc's Mechanics Dynamics Aesthetics (MDA) method \cite{Hunicke2004} and the USE model of user interaction with an automated system \cite{Cowley2008}.

At the highest level the model illustrates how a player's experience of the game arises from a player's participation in a game in several modes of interaction. A player's interaction with the underlying formal game system gives rise to a unique set of game play dynamics and a player receives either negative or positive feedback on performance etc. Thus one description of a game is as cybernetic system \cite{Salen2004}, i.e. a system with a control loop.

Adaptation as a part of this feedback loop can potentially provide improved control of the game system and thus a more tailored experience for individual players. This then gives rise to a more complex game dynamic and potential emergent game behaviour.

\subsection{Areas of player psychology to adapt to}
\label{sec:psych}

\subsubsection{Ability}
Ability within a game is influenced by a player's position within the \textit{learning curve} of the game. The learning curve is the sine qua non of game design: as argued by Koster \cite{Koster2005}, learning is the key ingredient that makes games fun.

Player ability is also influenced by their knowledge of the game's design patterns \cite{Bjork2005}. A player who is fluent with a particular design pattern, such as the 'Aim and Shoot' pattern, can have a higher skill level when beginning to play than a player who is not so familiar.

Learning and ability are related in information processing terms. In \cite{Cowley2008} it was pointed out that players process information from the game world, trying to balance the complexity of this environment with internal cognitive complexity. The complexity of the game comes from its control scheme, narrative, objectives, opponents and other such elements; while cognitive complexity refers to the player’s ability to take all that in and react, enabled through prior experience of the form or innate ability. So a balance between the two is desired. If the player cannot comprehend everything being thrown at him, he will be overwhelmed and unable to function in an ideal way. If the game does not provide sufficient challenge or interest, the player will be left unengaged and will lose motivation.

Thus we may say that negative imbalance leads to confusion and anxiety; and positive imbalance leads to boredom and apathy \cite{Rauterberg1995}. There is an echo of Ashby's Law of Requisite Variety in this formulation, since the variety needed to support learning and thus optimal game play must be present in both sides: player and game. In this sense, an adaptive single-player system would resemble a pair of linked homeostats \cite{Espejo1989}. On one side, the player learns the game system and/or narrative, attempting to 'beat' it by application of experience-based skill. On the other side, the game maintains its novelty by adapting to the player's current ability level. Therefore if one homeostat is a human player, the first requirement of an automated adaptive AI system is homeostasis of the player's experience. This means being able to keep up with the player's inevitable learning of the game system. Once that requirement is met, then the system can be tweaked to provide different levels of difficulty, types of experience, etc.

\subsubsection{Learning}
Learning is a key aspect of game play, and the fact of learning implies the necessity of some form of teaching. At the least, we can say that there is a didactic process inherent in the way game content is structured so that the player can learn it without being over- or underwhelmed. In a standard game, designing how this structure is revealed during play is the job of the game developer. In an adaptive game, the adaptive AI is forced to deal with player learning, perhaps by constraint to a given possibility space.

Controlling the pace of learning (or mastery) of players is integral to a game's design, as the quality of the play experience depends heavily upon it. Some games demand mastery with a levelling or 'power-up' structure in a very discrete, 'building block' way. Another class of games have learning built into the basic structure of game play. For instance Tetris or chess have relatively easy-to-learn mechanics but a great depth of emergent complexity, and the pace of learning follows the player’s own ability to uncover this complexity, enabled through practice.

DDA attempts to address these issues, but it must deal with a major hurdle: players vary in how much challenge they want to face, and DDA smoothes out the challenge. In other words, some players want to be challenged beyond their current abilities, and grow in skill to meet the challenge by replaying sections of the game over and over until they conquer the game. At the other end of the scale, some players want simply to wander, enjoy the game world and never be overly challenged, as discussed next.

\subsubsection{Personality \& interaction style}
The act of playing requires an attitude to the game being played that constitutes a personality, even in the case of AI agents where 'attitude' would only be attributable on observation by humans. The act of play requires commitment to a course of action that ends with an invested outcome, winning or losing being the most common type of outcome. The \textit{commitment} of a player, and their particular \textit{style} in undertaking the play actions, contribute to their play personality. We use commitment here to mean the dedication and steadiness a player shows in the act of playing. Style of play is used to encapsulate all the differences players may show in their approach to play tasks. As with any form of personality, a play personality is not to be thought of as static but quite contextualised and relative. It thus requires constant monitoring, with a dynamic player model.

Adapting play based on a fixed metric of player performance ignores the opportunity to refine adaptive AI based on types of players. This is a major problem with DDA, because adjusting difficulty adjusts the challenge of the game, and one difference between types (in many of the existing player typologies) is their preferred level of challenge. For instance, in the DGD typology \cite{Bateman2005}, the pure \textit{Conqueror} type requires very high challenge, while the pure Wanderer type requires stress-free play, i.e. little or no challenge. These are mutually exclusive and yet core requirements (of each type), so a game that ignores this in favour of adapting only to the player's evinced skill risks alienating both types. The subtle indicators of a player's type are in their \textit{approach} to play, not their skill in playing. Thus adapting play based on an evaluation of the player's type involves shifting the focus of play \textit{overall}, encompassing cohesive changes to difficulty, reward structure, aesthetics and automated assistance.

A player approaches a game from the unique perspective of her own play history and personality, as discussed above; in addition, players vary fundamentally in information processing styles. This has been addressed in the study of temperament theory, which is regarded as biologically based as personality is culturally based.
We, and others, have previously covered this topic in detail \cite{Cowley2016behavlet,Cowley2016:pre1,Bateman2011}. Thus it is sufficient to reiterate that there is a close link between the interaction styles that characterise people, and the patterns which reoccur throughout game design - and this is no accident because games are human artefacts designed for human minds.

\section{Solutions}
We can refer back to the idea of optimal experience in play as discussed in \cite{Cowley2008}. Adjusting the in-game elements measured is not done to 'put players in the Flow', but to make games reactive to individuals through tangible aspects of their experience for which affect is adjudged by heuristics derived from the Flow construct.

Referring to the model in figure \ref{fig.2pUSE}, adaptive AI involves leveraging information from the 'User(s)' to dynamically alter the elements of the central 'System' module, thus regulating activity within the right-hand 'Experience' module, which feeds back to the 'User(s)'.

Because this involves a feedback loop that achieves autonomy of control over elements of the game and consequently the play experience, it effectively takes over some control of that experience that was traditionally the sole preserve of the game designer. Nevertheless, adaptive AI should still be thought of as a just another tool of the game designer.

We suggest a two-pronged approach for game designers to effectively implement adaptive game AI: 
\begin{itemize}
	\item adapt a game at the mechanics layer – this is the level that fundamentally affects the game play. Adaptation at the mechanics layer impacts the dynamics of how the player plays and thus leads inevitably to a change at the aesthetic layer.
    
    \item give the formal specifications of in-game adaptive AI so that potential effects of their in-game adaptive methods on a player can be understood and plotted with some degree of accuracy.
    
\end{itemize}

\subsection{AI under constraint}
One key constraint for adaptive AI is that the player’s original conception of the game rules and elements, including \textit{a priori} knowledge \footnote{\textit{A priori} knowledge includes knowledge of 'realistic' or 'natural' elements. This can help when adapting, as some changes need not be explicitly explained, such as the trivial example of player opponents that increase in toughness as they increase in size. \textit{A priori} can also refer to game design patterns, existing conventions which somewhat binds developers to the forms of previous work in their chosen genre.}, must not change.

For instance, if the game is a simulation of competition or \textit{ag{\^o}n} \cite{Caillois1961} the player's opponents usually appear to have abilities similar to the player --- an example would be a fighting game like Street Fighter II (Capcom 1991). Adaptive AI should not suddenly change those abilities in an obvious way in mid-play, to adjust a game mechanic such as difficulty. Early examples of DDA in racing games caused dissonance among players by doing this: a terrible opponent who suddenly becomes lightning fast on the last lap would hurt players' immersion. 

In the DDA system for \textit{Max Payne} (3D Realms, 2001) was designed with this in mind, trying to make it invisible to players. The aim was to not be obvious when the game is self-adjusting its difficulty level, to maintain the game's immersion.

Yet hiding the rules in this way is a form of "black-box mechanics". Some game designers think this is bad practice, ergo the player should know about the adaptive elements, but we argue that there is always a complexity limit on players' knowledge of game mechanics. Thus they can be made aware of adaptive AI, if and only if they can be sufficiently informed of the logic under which the adaptive system works, so that they have some idea of why the game performs its actions. For instance, if the adaptive system is a non-linear and/or composed of complex rules or predicates, it may be excessively difficult to explain to the player. 

\subsubsection{Logical consistency}
The great thing in game design is to create a game with no \textit{capricious logic}. Capricious logic occurs when the game mechanics are not internally consistent, and this can occur for many reasons. A major cause is the fact that players \textit{observe} a game logic whose rules are often bent or broken for expediency or speed within the game engine.

That players demand a self-consistent logic from their games can be seen in Steinkuehler and Duncan's \cite{Steinkuehler2008} study of \textit{World of Warcraft} (Vivendi Universal 2004) players. They examined the cultural activities surrounding the game, such as online discussion forums. Here they discovered that players had been analysing game elements in an attempt to uncover hard information that would be useful in 'beating' the game. They claim that these are players of a young age with no scientific training, applying the scientific method to a game world because \textit{they trust that the internal logic of that world will be self-consistent}, so that applying logical analysis will bear fruit. The rule-based structure of games demand logic even if their playfulness should allow logic to be sometimes set aside

Logical self-consistency to a degree that would cope with this type of meta-gaming analysis could be considered a benchmark objective for in-game adaptive AI, since to adapt in real-time requires an autonomic element to the game engine. Pre-facing development of autonomous computer systems by modelling through formal methods should help to ensure their operational stability.

\subsubsection{Coherent 'Magic Circle'}
The second reason adaptive AI must be constrained is that players are temporarily redefining themselves and their world in terms of a new set of rules, defined by the game: a 'game world'. Huizinga \cite{Huizinga1949} describes play as a free and meaningful activity, carried out for its own sake, spatially and temporally segregated from the requirements of practical life, \textit{and bound by a self-contained system of rules that holds absolutely}. If these rules change players will be forced to 'step out' of the game to re-evaluate their perceived definition of the game world. This would destroy the sense of immersion in a game world which is important to the player's enjoyment of the 'fantasy' element of play.

This concept of a game world is known as the Magic Circle \cite{Huizinga1949} to games researchers. The Magic Circle pertains to the attitudinal psychology that is a pre-requisite of play, as individuals must take on the role of players in order to play. Huizinga held that the 'cheater' is less deleterious to other players' enjoyment than the 'spoil-sport', because the latter is denying the validity of the Magic Circle while the former is only trying to exploit it \cite{Swalwell2008}. The above example of \textit{World of Warcraft} players applying the scientific method can be thought of in the same way, since they are rejecting the attitude of playfulness in favour of production, sometimes known as 'the grind'.

\cite{Malone1980,Malone1981} also identified fantasy as an important part of enjoyment in gaming. Drawing on the psychology of intrinsic motivation, he was among the first to experiment on the relation between fantasy and game-play in educational games. The fact that fantasy is hugely motivating in game play is quite well-established now, and we are more interested in how that fantasy is structured. It is necessary to provide some comprehensible metaphor within the fantasy, so that players can easily digest the information content of the fantasy and go directly to dealing with the game mechanic. Thus, here again adaptive AI requires constraint.  

\paragraph*{Constraint summary}
So there is a potential conflict of interest between adaptive AI which can alter game mechanics, and preservation of logical consistency and the magic circle. For instance, one way to introduce novice players to a complicated control scheme is to begin with a restricted subset of the full scheme. But if this is how the player initially understands the game, they will question the introduction of new control dimensions unless they are explained within the narrative - i.e. in the acquisition of new equipment, skills, companions etc. Meeting the player's expectations for the logic and fantasy of their game is the key to creating effective adaptive components. These constraints on adaptive AI mean that great care must be taken when adapting in-game elements in real time - the player must either be forewarned that adaptation may take place (making it part of the game's rule set, which may conflict with realism), or must not be able to notice it at all.

\subsection{Abstraction}

One potential way to define adaptive AI constraints is by considering the game world abstraction. Games are necessarily an abstraction, because all games simulate some recognisable aspect of reality, be it an action, object or experience. Yet a game cannot be a facsimile simulation, because a facsimile implies reality with implied real consequences which are not game-like; so game designers abstract elements of their model of reality.

\begin{quote}
\textit{Game world abstractions state the rules of the world the game is set in: the nature of the game world, the player's potential interactions with that world, and the manner in which that world is represented }\cite{Bateman2005}.
\end{quote}

Game world abstractions also influence the design of game mechanics, as function follows form. How these mechanics can be adapted depends on the precise format for the abstraction, and how the player understands the abstraction. For example, if a player understands his flight simulator game to have high fidelity, by this abstraction and the logic that goes with it he will expect to be confronted by most of the stringent constraints of actual flight. Yet if mastery of this game is beyond him, logical consistency implies that he will not expect to suddenly be able to pull fantastical manoeuvres due to adaptive plane controls. Therefore, elements of the abstraction process can be examined for constraints \textit{and opportunities} for adaptation.

\subsection{Formal model}
One approach that can help to constrain the emergent qualities of adaptive AI is formal specification. Formal methods such as category theory \cite{Walters1991}, enable specification and verification of the objects and actions of the play space, and thus support rigorous testing of system coherence. While it is not a substitute for play testing, there are many advantages in testing algorithms and functions. Formal methods of category theory were applied to game specification in \cite{Grunvogel2005}.

Formal specification of the structure of game play should be built around our knowledge of axiomatic structures in the player-game interaction. For instance, we can treat the game as being a self-consistent structure of entities and (inter)action rules, which could be mapped to a grammar of Nouns and Verbs using natural (or more rarely, formal) language syntax. This is an intuitive description, as it comes close to what players see when beginning a game –-- objects/elements in the game which are classed as Nouns; and things Nouns can do which are classed as Verbs. The constraints above imply that adaptive AI can only alter the effect of a Verb or the makeup of a Noun, if the player knows a priori that it is a valid change or simply cannot observe it occurring.

Koster's \cite{Koster2005grammar} game description grammar deals with mechanics and somewhat with dynamics. The aim is to detail the algorithm that gives the possibility space of the game. The method notates how game play atoms would stack (in their hierarchy) and segue. The notation system includes rules or guidelines such as "atoms must have a failure state link" (which corresponds to game actions always having associated risk, i.e. being meaningful). An important rule of thumb is that atoms are structured around a core verb in the game play (like shoot, run or jump); and the hierarchy ends when that verb obtains certainty of outcome – i.e. the action loses meaning when the player loses freedom of choice.

This groundwork was built upon by Bura \cite{Bura2006}, who used a notation (derived from Petri-Nets) that specifies an isomorphism between the atomic transitions and the actions permitted by the operational rules of the game. Transitions in the nets represent discrete game state changes caused by a player's choice. Places in the net are relabelled resources, and can represent nouns and verbs (in the sense described above) - however \cite{Bura2006} gave an alternative formulation, in which resources can also represent abstract elements of game play, such as skill, preparation or luck. The Petri Net approach is taken further in \cite{Natkin2004} which uses extended hypergraphs to model together the game topology and the transaction net (which describes logical relationships).

\section{Discussion}
\label{sec:disc}
It is clear that adaptive AI can be enhanced using in-depth knowledge of the player in real-time. It is our primary argument in this paper that a generalised player model is the most efficient way to capture player individuality which is required for good function of adaptive AI. Properly constraining adaptive AI should therefore be a function of a player modelling approach. 

For instance, with a factor model of player characteristics we have an automated metric for the comprehensiveness of the player's internalisation of complex game components. Factor Modelling assigns values to the player's attributes (for instance \textit{Marksmanship}) in the game play structure, by utilising modelling techniques such as data clustering and descriptive Decision Theory. If the game developer had already implemented a formal grammar describing their game, its components would naturally lend themselves to a factor model – thus the proposed methods build on each other to give an integrated solution for constrained adaptive AI.

A similar but more comprehensive (and complex) approach is to use a formal specification applicable to interactive control systems, of which games are an example, and apply it to a generalised player model. In \cite{Cowley2016atari} the first author described such an approach. The formal specification was derived from the model for hybrid (discrete and continuous) control systems by \cite{Tabuada2004}, and applied to the Behavlets method for general player modelling \cite{Cowley2016behavlet}. The aim of this work is to represent Behavlets as action sequences in a formally defined simulation of a game system. The motivation is to generate a representation of possible player actions, and the archetypal behaviour traits that can shape those actions, such that the representation can be used as input for a machine learning system. Ultimately, the goal is to learn from real human behaviour.

The concept of Behavlets is to build a data-driven model of the player that is strongly linked to psychological theory. A formal specification serves to enable methods of comparison between abstract descriptions of gameplay, in addition to the aforementioned functions for creating constraints. Embedding Behavlets in a formal specification thus allows constrained adaptive gameplay systems which can be rigorously compared. Such comparison can help to design and tune the adaptive AI that a game designer has to create, supporting this paper's main question \textsf{RQ3}.

\subsection{Future work}
The earlier research questions \textsf{RQ1} and \textsf{RQ2} address the fundamentals of general player modelling. Satisfactory answers to these questions would enable the deployment of the same general player model to different games (at least of a similar type). This would then allow comparative study of multiple games. \textit{The same analytical approach will work for the same game under a finite set of adaptive AI conditions}. This is the core concept for our plan to address \textsf{RQ3}.

Thus the plan to specifically address \textsf{RQ3} specifies the use of Behavlets \cite{Cowley2016behavlet} to provide the generalised player model, and the formal method described in \cite{Cowley2016atari} to provide a rigorous framework for comparison. The method will involve iterative evaluation of progressively more complex adaptations of a testbed game such as a first-person shooter. The testbed game will be chosen to provide a well-understood experience, in the sense of being well-studied, and also a rich activity, in the sense of allowing players to express varying behaviours. \cite{Cowley2016behavlet} has guidelines on choosing a viable testbed game.

Thus, in the first instance the formally defined player model will be built for the testbed game with no adaptation conditions. The game will then be altered in a simple way, in order to demonstrate abstract game comparison-by-simulation \cite{Cowley2016atari}, comparing the original and altered versions. Thereafter, adaptive AI will be added to the modelled game, in multiple conditions with increasing complexity. Based on the generalised player model, each adaptation will be designed to respond to some aspect of the player profile, for example a player's tendency towards cautious play, creating multiple versions of the game. These versions will be evaluated by play testing to establish whether the generalised player model does in fact support adaptive AI. To further demonstrate that the system can facilitate adaptive AI, the versions with more complex adaptations will be compared with the simpler versions, using the simulation features of the formal model. The threshold for complexity will be set where it is no longer possible to manually test all possible adaptive outcomes. This will show that issues of logical consistency can be dealt with more readily when the profile of the player is known through a formal generalised player model.

\paragraph*{Issues}
The cited work, \cite{Cowley2016behavlet,Cowley2016atari}, is far from mature and like any fresh concept there are many details lacking. Behavlets as yet lack many features which could make them more efficient and easier to use, mainly by integrating machine learning. The state of the formal method presented may be incomplete, which may frustrate immediate attempts to apply it.
To bring these methods forward and provide a comprehensive answer for \textsf{RQ3} is a substantial piece of work. Ways to address specific issues of each method have been discussed in each publication. Ultimately, despite the seeming complexity, what is required to build such models is quite complementary to the development process - defining the entities and operations of game-play.

\subsection{Conclusion}
\label{sec:conc}
This preprint concludes our discussion of the background of generalised player models, and tasks required to build one. Addressing the three proposed RQs is suggested as a single program of work. To recap, work for \textsf{RQ1} will establish the state of the art for the generalised player model, covering the foundations such as methods for personality profiling as well as existing applications. Work for \textsf{RQ2} will focus on extending a computational intelligence-based player model with the features of a generalised player model. The work described above for \textsf{RQ3}, to improve adaptive game AI, will act as a validation of the whole approach in a scenario of interest to the game research and development communities alike.

Finally, following successful outcomes for \textsf{RQ1-3}, the methods used \cite{Cowley2016behavlet,Cowley2016atari} should be compared to competing generalised player modelling methods.

\newpage

\bibliographystyle{IEEEtran}
\bibliography{Cowley_Behavlet_preprint_3_bib}


\end{document}